\documentclass[prapplied,twocolumn,superscriptaddress,groupedaddress]{revtex4-1} 
\pdfoutput=1
\usepackage{graphicx}  
\usepackage{dcolumn}   
\usepackage{bm}        
\usepackage{amssymb}   
\usepackage[english]{babel} 
\usepackage{inputenc}
\usepackage{amsmath}
\usepackage{amssymb}
\usepackage{graphicx}
\usepackage{natbib}
\usepackage{mathrsfs}
\usepackage{color}

\begin{document}
\title{Propagation and imaging of mechanical waves in a highly-stressed single-mode phononic waveguide}

\author{E. Romero}
\thanks{These authors contributed equally to this work}
\author{R. Kalra}
\thanks{These authors contributed equally to this work}
\author{N. P. Mauranyapin}
\thanks{Corresponding author: n.mauranyapin@uq.edu.au}
\author{C. G. Baker}
\author{C. Meng}
\author{W. P. Bowen}
\affiliation{ARC Centre for Engineered Quantum Systems, School of Mathematics and Physics, The University of Queensland, Brisbane, Queensland 4072, Australia}

\begin{abstract}
We demonstrate a single-mode phononic waveguide that enables robust propagation of mechanical waves. The waveguide is a highly-stressed silicon nitride membrane that supports the propagation of out-of-plane modes. In direct analogy to rectangular microwave waveguides, there exists a band of frequencies over which only the fundamental mode is allowed to propagate, while multiple modes are supported at higher frequencies. We directly image the mode profiles using optical heterodyne vibration measurement, showing good agreement with theory. In the single-mode frequency band, we show low-loss propagation ($\sim1$~dB/cm) for a $\sim5$~MHz mechanical wave. This design is well suited for phononic circuits interconnecting elements such as non-linear resonators or optomechanical devices for signal processing, sensing or quantum technologies.	
\end{abstract}

\maketitle

\section{Introduction}

Advances in nano-electromechanical systems~\cite{ekinci_nanoelectromechanical_2005}, optomechanics~\cite{aspelmeyer_cavity_2014} and quantum computation~\cite{ladd_quantum_2010} have fuelled a resurgence in the field of phononic circuits. The field is motivated by the prospect to generate, route, interact and interface acoustic waves in on-chip architectures. While individual micro-mechanical devices have long been used for applications including signal processing~\cite{aigner_saw_2008}, frequency control~\cite{van_beek_review_2011} and sensing~\cite{bogue_recent_2013}, complex phononic circuits remain a challenge. Such circuits could perform efficient phononic information processing~\cite{sklan_splash_2015} with potentially low power consumption~\cite{roukes_mechanical_2004} and robustness in radiation harsh environments~\cite{chowdhury_single-device_2012}. Recent work has also explored phononic circuits in quantum technologies~\cite{ruskov_-chip_2013, habraken_continuous_2012,
vermersch_quantum_2017}, hybrid optomechanical systems~\cite{shin_control_2015,balram_coherent_2016} and topological metamaterials~\cite{cha_experimental_2018}.

Large-scale phononic circuits require that the acoustic waves be reliably guided between the various components in the circuit. Waveguides for electromagnetic waves at optical-, microwave- and radio-frequencies are mature technologies, having overcome issues of loss, scattering and dispersion. Phononic waveguides, on the other hand, are still an active area of research~\cite{hatanaka_phonon_2014,fang_optical_2016,
ghasemi_baboly_acoustic_2018,
liu_evanescent_2016}. The variety of channels through which mechanical vibrations can propagate represents a challenge for phononic waveguides, where imperfections can scatter energy into other modes. One key lesson from electromagnetic waveguides is that robust low-loss signal propagation is much easier to achieve when the waveguide is operated in a frequency band where only a single mode is available. This is evident in the use of the dominant TE$_{10}$ mode in rectangular microwave waveguides and in single-mode optical fibers which have enabled the long-distance communication that forms the backbone of the internet. Here, we translate this concept to phononics, demonstrating a low-loss single-mode phononic waveguide.

Recently, Patel et al.~\cite{patel_single-mode_2018} reported the first single-mode phononic waveguide, characterizing it through its coupling with an optomechanical resonator. The waveguide comprised of a nano-scale silicon beam supported by a phononic crystal, which serves to confine energy to the beam and modify its mode structure to create a single-mode frequency band. In contrast, we demonstrate single mode operation with a highly-stressed membrane which inherently eliminates all but out-of-plane modes in the frequency range of interest and enhances the quality factor that can be obtained. This design is motivated by increasing interest in developing phononic circuits based on interconnecting multiple micro- or nano-scale resonators~\cite{hatanaka_phonon_2014,
mahboob_interconnect-free_2011,
wenzler_nanomechanical_2014,
fon_complex_2017}. We characterize the waveguides using optical heterodyne vibration measurement to directly image the modes, demonstrating good agreement with theory. We find that, at room temperature and in vacuum, the waveguide permits low-loss propagation ($\sim 1$~dB/cm) in the single-mode regime.

\section{Theory}\label{sec:Theory}

\begin{figure}[t!]
\begin{center}
\includegraphics[width=\columnwidth]{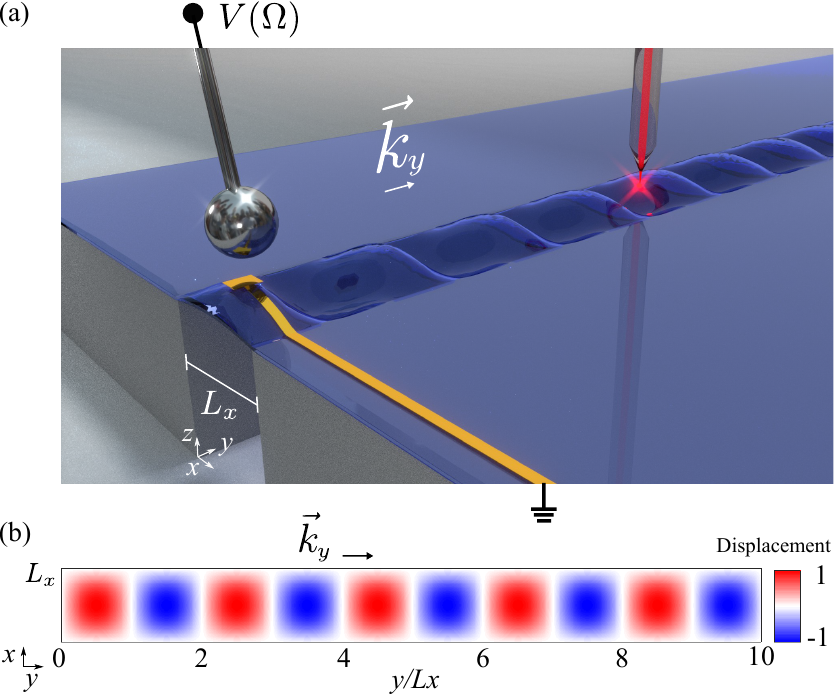}
\end{center}
\caption{(a) Illustration of the phononic waveguide with width $L_x$. The waveguide is actuated electrostatically through a capacitor formed by an electrode fabricated on the membrane (shown in gold) and a probe electrode (in silver). While a cross-section is shown here to highlight the membrane, the membrane is actually fixed on all sides by the substrate. Mechanical waves propagate along $y$ with wave vector $k_y$. Using a lensed optical fiber, the vibrations of the membrane are detected through their modulation of the phase of the laser light reflected back into the fiber. (b) Example of a snapshot of the calculated instantaneous out-of-plane displacement $\Re[u(x,y,t)]$ for a mode with $k_x=k_y=\pi/L_x$.}\label{fig:Diagram}
\end{figure}

We consider a waveguide made of a highly-stressed rectangular membrane with width $L_x$ that extends along the $y$-direction, the direction of propagation, as illustrated in Fig.\ref{fig:Diagram}(a). The acoustic impedance mismatch between the membrane and the supporting bulk material serves to confine the energy within the membrane~\cite{graczykowski_acoustic_2014}.

Vibrations of a membrane, where the restoring force is dominated by stress as opposed to the flexural rigidity of the material, satisfy the following wave equation~\cite{schmid_fundamentals_2016}:
\begin{equation}\label{eqn:Membrane_Wave}
\sigma \nabla^2 u(x,y,t)-\rho \frac{\partial^2}{\partial t^2}u(x,y,t)=0,
\end{equation}
where $u(x,y,t)$ is the displacement field along the out-of-plane direction $z$, $\sigma$ is the tensile stress of the membrane (N/m$^2$) and $\rho$ is the density of the membrane material (kg/m$^3$). The boundary conditions enforced by the clamping to the substrate are $u(0,y,t)=u(L_x,y,t)=0$.  The solution for the out-of-plane vibrations of the membrane driven at frequency $\Omega$ is~\cite{redwood_mechanical_1960}
\begin{equation}\label{eqn:Displacement_Field}
u(x,y,t)=\sum_{n} u_{n} \phi_{n}(x)\phi(y) e^{-i (\Omega t + \theta_n)},
\end{equation}
where $u_{n}$ and $\theta_n$ are the amplitude and phase of the $n^\mathrm{th}$ transverse mode, respectively, $\phi_{n}(x)= \sin(k_{x} x)$ is the transverse mode shape with $k_{x}= n \pi/L_x$, and $\phi(y)= e^{\pm i k_{y} y}$ is the propagating wave. The value of $k_y$ is obtained from the dispersion relation as derived from the wave equation,
\begin{equation}\label{eqn:Dispersion_Relation}
\Omega(k_{y})=\sqrt{\frac{\sigma}{\rho}}\sqrt{ k_{y}^2 + \left(\frac{n \pi}{L_x}\right)^2 }.
\end{equation}
We see from the dispersion relation that each transverse mode $n$ has a cut-off frequency $\Omega_{c,n}=\sqrt{\frac{\sigma}{\rho}}\left(\frac{n \pi}{L_x}\right)$ below which that mode cannot propagate. Thus, there is a frequency band over which the waveguide is single-mode, where $\Omega_{c,1} < \Omega < \Omega_{c,2}$. This is shown in Fig.~\ref{fig:Modeshape_Spectra}(a), where we plot the dispersion relation calculated for a silicon nitride membrane with width $L_x=76~\mu$m, thickness $h=80$~nm and internal tensile stress $\sigma=1~$GPa. Highlighted in grey is the frequency band where there are no modes available for propagation, while the single-mode region is highlighted in blue. 

In practice, we measure a waveguide with finite length $L_y$, imposing the boundary conditions $u(x,0,t)= u(x,L_y,t)=0$. This transforms the waveguide into a resonator with standing waves $\phi_{m}(y)= \sin(k_{y} y)$, where $k_{y}= m \pi/L_y$ and $m$ is the longitudinal mode number. In Fig.~\ref{fig:Diagram}(b), we show an example of the normalized instantaneous displacement $\Re[u(x,y,t)]$ for the case $k_x=k_y$ and $L_y = 10L_x$.

\section{Experimental Methods}

\subsection{Fabrication}\label{sec:Fabrication}

The phononic waveguides are fabricated from high-stress ($\sigma\sim$1~GPa) low-pressure chemical vapour deposited Si$_3$N$_4$ films deposited on 500~$\mu$m Si wafers. They have dimensions $L_x=75$-$80~\mu$m, $L_y\approx1.8~$cm and thickness $h=80$~nm. The fabrication process is illustrated in Fig.~\ref{fig:Experimental}(a), with steps labelled (i-vi). (i) The process begins with a silicon wafer double-side coated with a silicon nitride thin film. (ii) Gold electrodes are patterned on the top side of the wafer using negative photoresist and a lift-off process. (iii) Both the top and bottom surfaces of the wafer are coated with 150~nm of aluminium followed by photoresist. The aluminium on the bottom side is patterned with photolithography and wet etched to serve as an etch mask. The photoresist and aluminium on the top side will protect the surface during bottom side patterning and etching, respectively. (iv) The exposed silicon nitride is etched using reactive ion etching (RIE) combining SF$_6$ and CF$_4$ gases. (v) The patterned aluminium is used as a mask for back side etching of roughly $480~\mu$m of silicon using deep reactive ion etching (DRIE). (vi) The final release of the silicon nitride membrane and aluminium removal is done using a highly concentrated KOH solution. The KOH etch of silicon is anisotropic and forms angled end facets following the Si(111) plane. This process straightens the edges of the waveguide in a self-limiting process (see Fig.~\ref{fig:Experimental}(c,d)).

\subsection{Measurement Setup}\label{sec:Detection}

Measurements are performed at room temperature with the waveguide in a vacuum chamber with pressure $\sim 10^{-7}$~mBar~\cite{naesby_effects_2017}. The waveguide is actuated electrostatically through a capacitor formed by the on-chip electrode and a probe electrode held $\sim5$~$\mu$m above the waveguide, as shown in Fig.~\ref{fig:Diagram}(a). The force applied is proportional to $(V_\mathrm{DC}+V_\mathrm{AC}\cos{\Omega t})^2$, where $V_\mathrm{DC}$ is the applied DC bias and $V_\mathrm{AC}$ is the amplitude of the AC drive at frequency $\Omega$~\cite{bekker_injection_2017}. To measure the vibrations on the waveguide, we use an optical interferometric heterodyne detection scheme, represented in Fig.~\ref{fig:Experimental}(b). A Ti-Sapphire laser beam with wavelength $\lambda=780~$nm is passed through an acousto-optical modulator (AOM) driven at 77~MHz to obtain a frequency-upshifted local oscillator beam as well as a non-diffracted probe beam at 780nm. The probe beam is focused onto the membrane through a lensed fiber with spot size $\sim 1~\mu$m (Fig.\ref{fig:Diagram}(a)). The membrane vibrations modulate the phase of the probe light reflected back into the lensed fiber, which interferes with the local oscillator to create beat notes at $\omega_\mathrm{AOM} \pm \Omega$. With low-amplitude vibrations, the amplitude of the photocurrent at frequency $\omega_\mathrm{AOM} \pm \Omega$ is, to a good approximation, directly proportional to the amplitude of vibration of the membrane $\left|u(x,y)\right|$. We use a spectrum analyser for measurement of the photocurrent at frequency $\omega_\mathrm{AOM} + \Omega$ to obtain the mechanical response. Both the electrode and lensed fibre are mounted on piezoelectric positioners to allow them to be scanned during measurement.

\begin{widetext}
\onecolumngrid
\begin{figure}[t]
\includegraphics[width=\columnwidth]{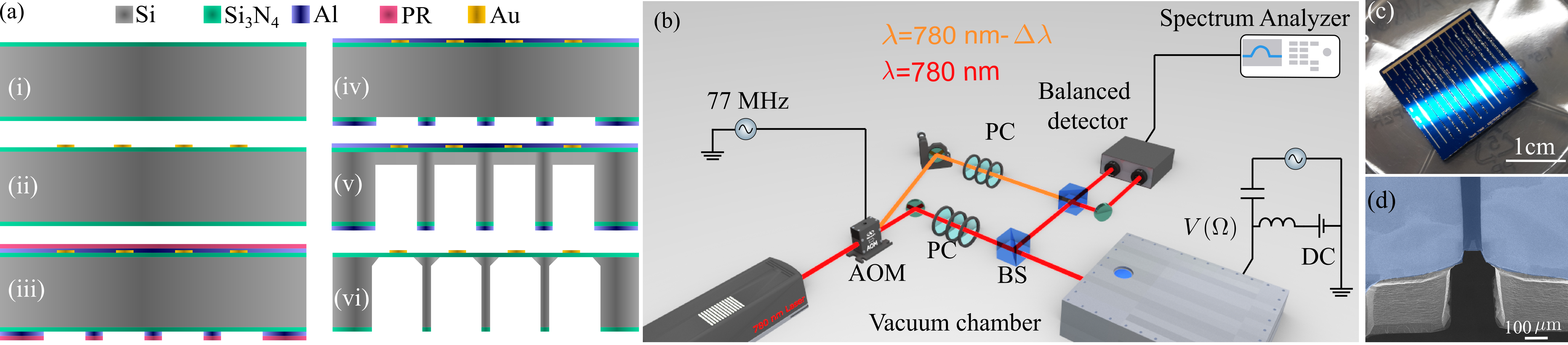}
\caption{(a) Fabrication steps for the phononic waveguides (see main text for details). (b) Schematic of the optical heterodyne detection setup, complete with an acousto-optical modulator (AOM), polarization controllers (PC) and beam splitters (BS), used to measure the mechanical vibrations of the waveguide. The diffracted local oscillator (orange) and non-diffracted probe beams are combined in a final beam splitter for balanced photo-detection. The voltage $V(\Omega)$ required for actuation is delivered to the waveguide in the vacuum chamber via a feed-through. (c) Image of a chip with 14 waveguides in parallel, four of which are intact. (d) False color cross-sectional scanning electron micrograph of the released Si$_3$N$_4$ membrane (blue) held by the Si substrate (grey).}\label{fig:Experimental}
\end{figure}
\twocolumngrid
\end{widetext}

\section{Measurement Results}

\subsection{Mode Shape Imaging}

\begin{figure}[t]
	\includegraphics[width=\columnwidth]{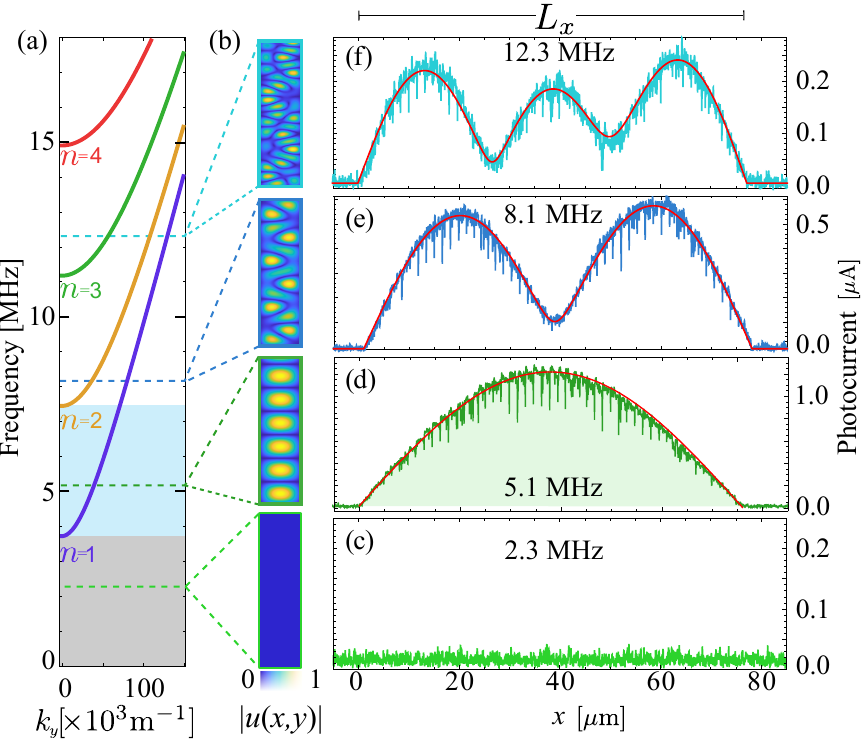}
	\caption{(a) Calculated dispersion relation for the waveguide, showing the first four mode branches. Shaded in blue and gray are the frequency bands corresponding to the single-mode region and the band below cut-off, respectively. (b) Examples of calculated two-dimensional mode profiles $\left|u(x,y)\right|$ at the frequencies indicated by dashed lines in (a), highlighting the interference between modes with different longitudinal wavevectors in the multi-mode bands. (c-f) Experimentaly measured transverse mode profiles of the phononic waveguide when actuated at $\Omega/2\pi=$ (c) 2.3~MHz, (d) 5.1~MHz, (e) 8.1~MHz and (f) 12.3~MHz. The photocurrent at frequency $\omega_\mathrm{AOM} + \Omega$ is plotted as a function of the $x$ position of the fiber while it is scanned across the waveguide of width $L_x \approx 76~\mu$m. The red curves are fits to the data (see main text for details). The fiber is positioned approximately 3~mm from the actuation electrodes along $y$, the direction of propagation. As indicated by the dashed lines in (a), the measurements correspond to the (c) zero-, (d) one-, (e) two- and (f) three-mode frequency bands.}\label{fig:Modeshape_Spectra}
\end{figure}

In a first experiment, we image the transverse mode shape while the membrane is continuously actuated at a frequency $\Omega$. This is repeated at different frequencies lying within the bands below cut-off, in the single-mode region and in the multi-mode region, as marked by the dashed lines in the plot of the dispersion relationship, Fig.~\ref{fig:Modeshape_Spectra}(a). Two scenarios are possible. Given the finite length of the waveguide ($L_y \approx 1.8$~cm), standing waves will be generated if the propagation and reflection losses are sufficiently low. If the losses are high, however, propagating waves would be observed which are not significantly perturbed by reflections at the end of the waveguide. While the two cases would result in similar profiles, we know from experiments presented in the next section that standing waves are generated. We will thus restrict our discussion of mode profiles to those corresponding to standing waves.

Fig.~\ref{fig:Modeshape_Spectra}(b) shows examples of the calculated two-dimensional profiles at the chosen frequencies, where oscillation amplitude $\left|u(x,y)\right|$ is plotted as a function of $x$ and $y$ along the waveguide. In contrast to the simple profile expected in the single-mode regime, the profiles at higher frequencies are a result of the interference between multiple modes and depend both on their relative amplitudes and phases with respect to the drive tone.

To image the transverse mode shape, the photocurrent at frequency $\omega_\mathrm{AOM} + \Omega$ is recorded while the lensed fibre is scanned over the width of the waveguide. The scan has an average speed of $1~\mu$m/s with a 200~nm step size. Fig.~\ref{fig:Modeshape_Spectra}(c,d,e,f) show the measured profiles for actuation at $\Omega/2\pi=2.3$, 5.1, 8.1 and 12.3~MHz, respectively. As expected, the waveguide does not respond when driven at 2.3~MHz, well below the expected cut-off frequency of 3.7~MHz, as shown in Fig.~\ref{fig:Modeshape_Spectra}(c). Driving in the single-mode frequency band results in the expected $\sin(\pi x/L_x)$ shape of the fundamental transverse mode, as shown in Fig.~\ref{fig:Modeshape_Spectra}(d). The measured photocurrent amplitude of approximately 1~$\mu$A corresponds to an actual vibration amplitude of $\sim10$~pm. 

At 8.1~MHz and 12.3~MHz, the waveguide supports two and three modes, respectively. The two and three anti-node character of the $n=2$ and $n=3$ modes are clearly visible in Fig.~\ref{fig:Modeshape_Spectra}(e) and (f). To obtain fits to the data, we tune the relative amplitudes and phases of each standing wave. For example, at $\Omega/2\pi = 8.1$~MHz, the $n=1$ and 2 are expected to have longitudinal wavelengths $\lambda_{y,1}=2\pi/k_{y,1}= 77~\mu$m and $\lambda_{y,2}=2\pi/k_{y,2}= 169~\mu$m, respectively. The fit is obtained with a relative phase difference of $\theta_2 - \theta_1 = 1.3$~rad and $u_2/u_1 = 1.8$ at a certain point along the waveguide.

\begin{figure}[t]
	\includegraphics[width=\columnwidth]{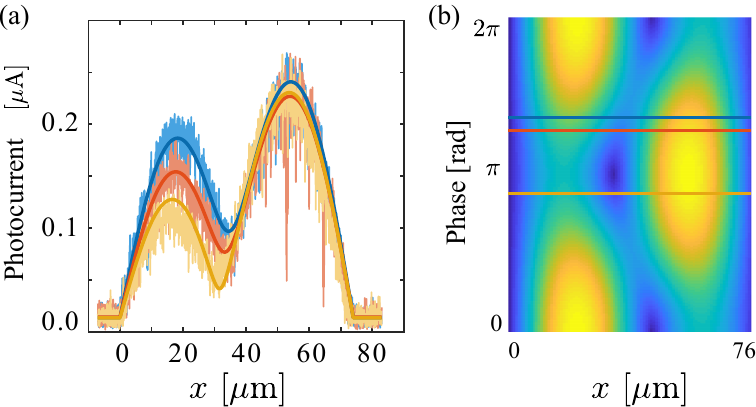}
	\caption{(a) Consecutive measurements of the transverse mode profile of the waveguide actuated at a fixed frequency $\Omega/2\pi = 10.12$~MHz and at a fixed $y$ position. The changing mode profiles show that the resonances forming the standing waves measured experience drift over the timescale of minutes. (b) Calculated transverse mode profile $\left|u(x,y)\right|$ as a function of the relative phase between the $n=1$ and 2 modes at a certain distance from the edge of the waveguide with $u_2/u_1 = 3.3$. The fits in (a) are taken from slices of this plot as labelled, supporting the hypothesis of a phase drift.}\label{fig:Interference_Modes}
\end{figure}

While the measurements shown in Fig.~\ref{fig:Modeshape_Spectra} fit well with the mode shapes we would expect from perfect standing waves, we do observe that the relative phase drifts on the time-scale of minutes. To show this, we drive the waveguide at frequency $\Omega/2\pi = 10.2$~MHz, exciting the $n = 1$ and $n = 2$ modes. We measure the transverse mode shape as done before, repeating this over the span of a few minutes at the same $y$ position. Three consecutive measurements are plotted in Fig.~\ref{fig:Interference_Modes}(a), demonstrating that the measured mode shape clearly changes over time. This can be attributed to a drift of the relative phases of the resonances with respect to the drive tone. Fig.~\ref{fig:Interference_Modes}(b) shows calculations of the expected profile as a function of the relative phase between the $n = 1$ and $n= 2$ modes. The fits to the data in Fig.~\ref{fig:Interference_Modes}(a) are obtained from the calculations shown in Fig.~\ref{fig:Interference_Modes}(b) at different relative phases, as labelled in the figure, supporting the hypothesis of a slow phase drift.

\subsection{Frequency Response}

To characterize the frequency response of the phononic waveguide, we perform a network analysis with the fiber positioned at the center of the waveguide, $x=38~\mu$m, and at $y=3$~mm as before. The drive frequency is swept while the maximum response at each frequency is captured on the spectrum analyzer. We observe that the mechanical response is indeed suppressed below the cut-off frequency. The response above cut-off is populated with resonant peaks which drift on the time-scale of minutes, consistent with observations presented in Fig.~\ref{fig:Interference_Modes}. To capture the linewidth of an individual resonance, we repeat the measurement over a narrow frequency window and over a time short compared to the drift. Fig.~\ref{fig:Network_analyzer}(a) shows a measured resonance at $\Omega/2\pi = 4.754$~MHz, which is in the single-mode frequency band. Fitting this peak to a Lorentzian, whose symmetry is broken by the presence of a resonance at lower frequency, we estimate a quality factor of $Q= 2100\pm 200$.

\subsection{Pulsed Actuation}\label{sec:Pulsed}

In order to provide another measure of the quality factor of the resonance at 4.754~MHz, we perform ring-up and ring-down measurements. With the fibre positioned 8.8~mm away from the actuation electrodes, we gate the drive tone with a square pulse with a width chosen to be longer than the expected ring-up time of the cavity. In order to resolve these dynamics we acquire data with a resolution bandwidth ranging from 100~kHz to 1~MHz, corresponding to a  integration time of 1~$\mu$s to 10~$\mu$s. To improve the signal-to-noise ratio, we average over 2000 repetitions synchronized to an external trigger.

\begin{figure}[t]
	\begin{center}
		\includegraphics[width=\columnwidth]{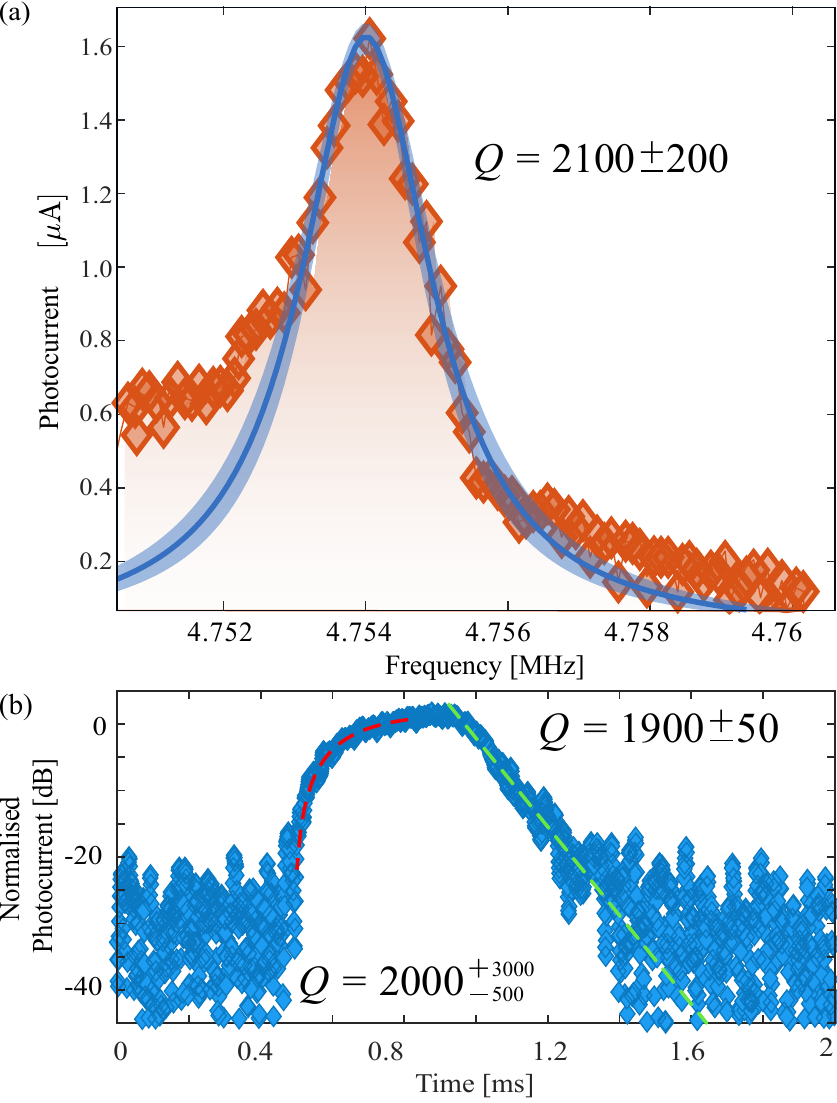}
	\end{center}
	\caption{(a) Frequency response of the waveguide highlighting a resonance at $\Omega/2\pi=4.754$~MHz. The peak is fit to a Lorentzian (in blue with the shading denoting uncertainty) from which we extract a quality factor $Q= 2100 \pm 200$. (b) Response of the waveguide to a 5~ms pulse at 4.754~MHz. The normalized photocurrent is plotted in decibels from which the ring-up and ring-down times are extracted, yielding quality factors of $Q= 2000^{+3000}_{-500} $ and $1950\pm 50$, respectively.}\label{fig:Network_analyzer}
\end{figure}

Fig.~\ref{fig:Network_analyzer}(b) shows the normalized photocurrent as a function of time. The ring-up at the start of the pulse is fitted to the expected exponential build-up for a simple-harmonic oscillator, from which we extract a quality factor of $Q= 2000^{+3000}_{-500} $. The ring-down at the end of the pulse is fitted to an exponential decay which yields $Q =1950 \pm 50$, in fair agreement with our other estimates of $Q$. We can translate the measured quality factors to a conservative estimate of the propagation loss $\alpha$ along the waveguide. With $Q\approx1900$, we estimate $\alpha\approx 1$~dB/cm. This is conservative since it neglects the fact that imperfect reflections at the ends of the cavity can also contribute to the measured decay rate of the cavity. Nevertheless, it compares favourably to previous reports of room temperature megahertz waveguides~\cite{hatanaka_phonon_2014}.

\vspace{0.5cm}

\section{Conclusion}

In this work we have demonstrated a single-mode phononic waveguide based on a highly-stressed silicon nitride membrane. Direct imaging of the mode shapes in the single- and multi- mode regions show good agreement with the analytical solutions of the system. The finite-length waveguide we measure has sufficiently low propagation loss and reflection from the end surfaces to behave as a resonator with a quality factor $\sim 2000$ in the single-mode region. This provides a conservative estimate of the propagation loss: $\sim 1$~dB/cm, more than an order of magnitude lower than previous room-temperature phononic waveguides operating at megahertz frequencies~\cite{hatanaka_phonon_2014}. 

The waveguide is the first step towards developing scalable phononic circuits, which may find application in sensing~\cite{coulombe_computing_2017}, computation~\cite{roukes_mechanical_2004} and even quantum technologies~\cite{ruskov_-chip_2013, habraken_continuous_2012,
vermersch_quantum_2017,
shin_control_2015,balram_coherent_2016}. The structural and mathematical similarity of our implementation to microwave and optical waveguides paves a direct path towards developing other circuit components. It is possible to encourage, in future, complex phononic circuits built from phononic waveguides, processing and communicating information, interconnecting nanomechanical sensors, interfacing to photonic and microwave systems and, generally, providing new capabilities for on chip classical and quantum information systems.

\section{Acknowledgements}

This research was primarily funded by the Australian Research Council and Lockheed
Martin Corporation through the Australian Research Council Linkage Grant LP160101616. Support was also provided by the Australian Research Council Centre of Research Excellence for Engineered Quantum
Systems (CE170100009). R.K., C.G.B and W.P.B. acknowledge fellowships from the University of Queensland (UQFEL1719237 and UQFEL1833877) and the Australian Research Council (FT140100650), respectively. This work was performed in part at the Queensland node of the Australian National Fabrication Facility, a company established under the National Collaborative Research Infrastructure Strategy to provide nano and microfabrication facilities for Australia's researchers. The authors thank Luke Uribarri, George Brawley and Hafiz Abdullah for useful discussions, and Doug Mair and Kai-Yu Liu for technical assistance.

\bibliography{PRApplied_Library}

\end{document}